\documentclass[prb,twocolumn,showpacs,amsmath,amssymb,superscriptaddress]{revtex4}
\usepackage[english]{babel}
\usepackage{graphicx}
\newcommand{\lsim}{\lesssim}
\newcommand{\gsim}{\gtrsim}
\begin{document}
\title{Dynamic Spin Structure Factor of SrCu$_{2}$(BO$_{3}$)$_{2}$ at Finite Temperatures}
\author{S. \surname{El Shawish}}
\affiliation{J. Stefan Institute, SI-1000 Ljubljana, Slovenia}
\author{J. \surname{Bon\v ca}}
\affiliation{J. Stefan Institute, SI-1000 Ljubljana, Slovenia}
\affiliation{Faculty of Mathematics and Physics, University of Ljubljana, SI-1000 Ljubljana, Slovenia}
%
%
\author{I. \surname{Sega}}
\affiliation{J. Stefan Institute, SI-1000 Ljubljana, Slovenia}

\date{\today}
\begin{abstract}
Using finite temperature Lanczos technique on finite clusters we
calculate dynamical spin structure factor of the quasi-two-dimensional
dimer spin liquid SrCu$_{2}$(BO$_{3}$)$_{2}$ as a function of
wavevector and temperature. Unusual temperature dependence of
calculated spectra is in agreement with inelastic neutron scattering
measurements. Normalized peak intensities of the single-triplet peak
are ${\bf q}$-independent, their unusual temperature dependence is
analyzed in terms of thermodynamic quantities.
\end{abstract}
\maketitle

\section{Introduction}
In low-dimensional quantum spin systems quantum fluctuations often
lead to disordered ground states that exhibit no magnetic ordering and
a gapped, non-degenerate singlet ground state. Such states, also
called spin liquids, are realized in one dimension in dimerised or
frustrated spin chains, even-leg spin ladders as well as in the two
dimensional Shastry-Sutherland (SHS) model.\cite{shastry81}
SrCu$_{2}$(BO$_{3}$)$_{2}$ is a quasi-two-dimensional spin system with
a unique spin-rotation invariant exchange topology that leads to a
singlet dimer ground state. \cite{smith91} Since this compound
represents the only known realization of the SHS model, it recently
became a focal point of theoretical as well as experimental
investigations in the field of frustrated spin systems. Consequently,
many fascinating physical properties of SrCu$_{2}$(BO$_{3}$)$_{2}$
have been discovered. Increasing external magnetic field leads to a
formation of magnetization plateaus \cite{kageyama99,onizuka00} which
are a consequence of repulsive interaction between almost localized
triplets. Weak anisotropic spin interactions can have
disproportionately strong effect in a frustrated system.  It has
recently been shown, that the inclusion of the nearest neighbor (NN)
and next-nearest neighbor (NNN) Dzyaloshinsky-Moriya (DM) interactions
is required to explain some qualitative features of the specific heat
near the transition from the spin dimer to the spin-triplet state, as
well as to explain the low frequency lines observed in electron spin
resonance experiments in SrCu$_{2}$(BO$_{3}$)$_{2}$.
\cite{cepas01,cepas02,nojiri03,jorge03,zorko00,zorko01,elshawish}

The existence of the spin gap, almost localized spin-triplet excited
states, as well as the proximity of a spin-liquid ground state of
SrCu$_{2}$(BO$_{3}$)$_{2}$ to the ordered antiferromagnetic state,
lead to rather unusual low-temperature properties emerging in
inelastic neutron scattering (INS),\cite{kageyama00,gaulin} Raman
scattering (RS) \cite{lemmens} as well as in electron spin resonance
(ESR) experiments. \cite{nojiri03} In particular, INS normalized peak
intensities of single-, double- and possibly triple-modes show a rapid
decrease with temperature around 13~K, well below the value of the
spin gap energy $\Delta \sim 34$~K.  In addition, authors of
Ref.~\cite{gaulin} show, that properly normalized complement of static
uniform spin susceptibility, obtained with almost identical
model parameters as in the present work,\cite{jorge03} nearly
perfectly fits their experimental data. Similar behavior is found in
RS data where a dramatic decrease of Raman modes, representing
transitions between the ground state and excited singlets, with
increasing temperature at $T\ll \Delta$ is observed.  Moreover, at
$T\sim \Delta$ all RS modes become strongly overdamped.\cite{lemmens}

Numerical simulations of dynamical spin structure factor based on
exact diagonalization on small clusters at zero temperature show good
agreement with INS data.\cite{miyahara03} Recently developed
zero-temperature method based on perturbative continuous unitary
transformations \cite{knetter,knetter1} gives very reliable results
for the dynamical spin structure factor of the SHS model since the
method does not suffer from finite-size effects. The method is,
however, limited to calculations at zero temperature and, at least at
the present stage, it does not allow for the inclusion of DM terms.

The aim of this work is to investigate finite temperature properties
of the dynamical spin structure factor of the SHS model using the
finite temperature Lanczos method (FTLM),\cite{jj1,jj2} and to compare
results with INS data.\cite{kageyama00,gaulin} In our search for
deeper physical understanding of spectral properties of the SHS model
at finite temperatures we compare those with thermodynamic quantities,
such as: the specific heat, entropy, and uniform static magnetic
susceptibility, which we further compare with analytical results of
the isolated dimer (DIM) model. We finally present results of the
${\bf q}$-dependent static magnetic susceptibility as a function of
$T$.

\section{Model}

To describe the low-temperature properties of
SrCu$_{2}$(BO$_{3}$)$_{2}$ we consider the following Heisenberg
Hamiltonian defined on a 2D Shastry-Sutherland lattice:
\cite{shastry81}
\begin{eqnarray}
\nonumber
H_s\!\! &=&\!\! J \sum_{\langle {\bf i, j} \rangle}  {\bf S_{i} \cdot S_{j}}
+ J' \sum_{\langle {\bf i, j} \rangle'} {\bf S_{i} \cdot S_{j}} \\
&\,+&\!\! \sum_{\langle {\bf i \rightarrow j} \rangle'} {\bf D'} \cdot
({\bf S_{i} \times S_{j}}).
\label{Hamil}
\end{eqnarray}
Here, $\langle {{\bf i,j}} \rangle$ and $\langle {{\bf i,j}} \rangle'$
indicate that ${\bf i}$ and ${\bf j}$ are NN and NNN,
respectively. 
A recently published high-resolution INS measurments on
SrCu$_{2}$(BO$_{3}$)$_{2}$ \cite{gaulin} motivated us to choose
$J=76.8$~K (in units of $k_B$) and $J'/J=0.62$. A slightly different
choice of parameters ($J=74.0$~K, $J'/J=0.62$) has been used
previously in describing specific heat measurements \cite{jorge03} and
ESR experiments on SrCu$_{2}$(BO$_{3}$)$_{2}$.\cite{elshawish} We
should note, however, that this fine-tunning of parameters leads to
effects, visible only on small energy scales, thus leaving previous
calculations \cite{jorge03,elshawish} practically unaffected.
In addition to the Shastry-Sutherland Hamiltonian, $H_s$ includes DM
interactions to NNN with the corresponding DM vector ${\bf D'}$. The
arrows indicate that bonds have a particular orientation as described
in Ref. \cite{jorge03}.  Its value, ${\bf D'}=1.77~{\rm K}~{\hat{\bf
z}}$, successfully explains the splitting between the two
single-triplet excitations observed with ESR \cite{cepas01,nojiri03}
and INS measurements.\cite{kageyama00,gaulin}


As pointed out in Refs.~\cite{jorge03,elshawish}, a finite NN DM term
should also be taken into account to explain specific heat data and
ESR experiments. We have chosen to omit this term since it does not
significantly affect results of the dynamical spin structure factor at
a non-zero value of the wavevector.  We have chosen the quantization
axis $\hat{\bf z}$ to be parallel to the $c$-axis and $\hat{\bf x}$ to
the $a$-axis pointing along the centers of neighboring parallel
dimers.

We use the FTLM based on the Lanczos procedure of exact
diagonalization, combined with random sampling over initial wave
functions. For a detailed explaination of the method and a definition
of method parameters see Refs. \cite{jj1,jj2}. All the results are
computed on a tilted square lattice of $N=20$ sites with $M_1=100$
first and $M_2=250$ second Lanczos steps, respectively. The full trace
summation over $N_{st}=2^N$ states is replaced by a much smaller set
of $R\sim 500$ random states giving the sampling ratio $R/N_{st}\sim
5\cdot 10^{-4}$.
Comparing FTLM with the conventional Quantum Monte Carlo (QMC) methods
we emphasize the following advantages: (a) the absence of the
minus-sign problem that usually hinders QMC calculations of frustrated
spin systems, (b) the method connects the high- and low-temperature
regimes in a continuous fashion, and (c) dynamic properties can be
calculated straightforwardly in the real time in contrast to employing
the analytical continuation from the imaginary time, necessary when
using QMC calculations. Among the shortcomings of FTLM is its
limitation to small lattices that leads to the appearance of
finite-size effects as the temperature is lowered below a certain
$T<T_{fs}$. Due to the existence of the gap in the excitation spectrum
and the almost localized nature of the lowest triplet excitation, we
estimate $T_{fs}\sim 1$~K at least for calculation of thermodynamic
properties.\cite{jorge03} Finite-size effects also affect dynamical
properties as, e.g., the dynamical spin structure factor, which is
(even at finite temperatures) represented as a set of delta
functions. In particular, finite-size effects affect the frequency
resolution at low temperatures, while at higher temperatures as more
states contribute to the spectra, its shape becomes less size
dependent.

We should stress that FTLM was in the past successfully used in
obtaining thermodynamic as well as dynamic properties of different
correlated models as are: the $t$-$J$ model,\cite{jj1,jj2} the Hubbard
model,\cite{bonca02} as well as the SHS model.\cite{jorge03}

\section{Numerical Results and Comparison With Experiment}
\subsection{Dynamical Spin Structure Factor}
For comparison with the INS data, we compute the dynamical spin
structure factor
for $\mu=x,y,z$
\begin{eqnarray}
S_{\mu\mu}({\bf q},\omega)\!\!&=&\!\! {\rm Re}\int\limits_0^\infty\!\!dt\ e^{i\omega t}
\left\langle S^\mu_{\bf q}(t)S^\mu_{\bf-q}(0)\right\rangle;\nonumber \\
S^\mu_{\bf q}\!\!&=&\!\! \frac{1}{\sqrt{N}}\sum_{i,\alpha}S^\mu\left({\bf R}_i+ {\bf
r}_\alpha\right)e^{i{\bf q}\left({\bf R}_i+{\bf r}_\alpha\right)},
\label{som}
\end{eqnarray}
where $i$ runs over all unit cells of the lattice and ${\bf r}_\alpha$ spans
four vectors forming the basis of the unit cell that contains two
orthogonal dimers. For the details describing interatomic distances we
refer the reader to Ref.~\cite{knetter}.  The average in
Eq.~(\ref{som}) represents the thermodynamic average which is computed
using FTLM.\cite{jj1}

\begin{figure}[htb]
\includegraphics[angle=0,width=7.2cm,scale=1.0]{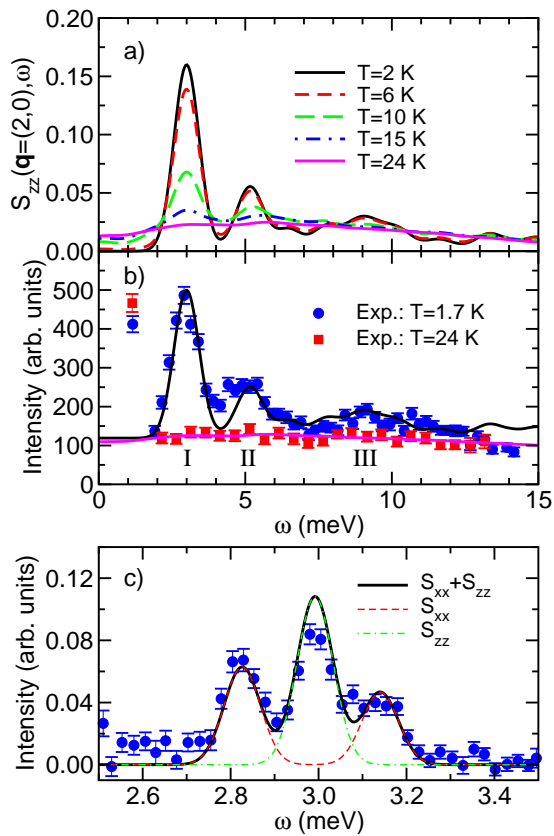}
\vspace{0.4cm} \caption{(Color online) (a) Spin structure factor $S_{zz}({\bf
q},\omega)$ for ${\bf q}=(2,0)$ where ${\bf q}$ is given in units of
the reciprocal lattice vectors \cite{defq} vs. $\omega$ for different
values of $T$, and (b) comparison with INS measurements from
Ref.~\cite{kageyama00}. Transitions to single-, double-, and
triple-triplet states are indicated with Roman numerals.
(c) Sum of the transverse $S_{yy}({\bf q},\omega)$ and longitudinal
$S_{zz}({\bf q},\omega)$ contributions of a single-triplet mode
compared to the high-resolution data for ${\bf q}=(-2,0)$ and
$T=1.4$~K from Ref.~\cite{gaulin}. We use units where $\hbar =1$.}
\label{fig1}
\end{figure}

In Fig.~\ref{fig1}(a) we first present the spin structure factor
$S_{zz}({\bf q},\omega)$ for different temperatures. We should note
that due to a finite value $D'_z = 1.77$~K spin rotational invariance
of the Hamiltonian, Eq.~(\ref{Hamil}), is broken,
{\it i.e.}  $S_{xx}({\bf q},\omega)=S_{yy}({\bf q},\omega)\ne
S_{zz}({\bf q},\omega)$.
Since $D'_z\ll J,J^\prime$, the effect of broken symmetry is, at least
within our numerical precision, negligible
for energy resolutions much larger than the value of anisotropic
interaction, $\Delta\omega\gg D'_z$,
yielding nearly identical results for the three components $\mu=x,y,$
and $z$ of $S_{\mu\mu}({\bf q},\omega)$. Since the spectra consist of
a set of delta functions, we have artificially broadened the peaks
with a Gaussian form with $\sigma=0.4$~meV, to achieve the best fit
with INS measurements.\cite{kageyama00} Two peaks (I and II) are
clearly visible at low temperatures $T=2~{\rm K}\ll \Delta$ around
$\omega\sim 3$~meV and $5$~meV, associated with transitions to single-
and double-triplet states. A broader peak (III) around $\omega=9$~meV
represents transitions to triple-triplet states.  Results at $T=2$~K
are consistent with previous $T=0$ simulations.
\cite{miyahara03,knetter}  Increasing temperature has a pronounced
effect on $S_{zz}({\bf q},\omega)$, manifesting in a rapid decrease
with temperature, at temperatures even far below the value of the gap.
Quantitatively, at $T=24$~K the peak structure almost completely
disappears. In Fig.~\ref{fig1}(b) we present comparison of our
numerical data scaled and shifted along the vertical axis to
compensate for experimental background for two different temperatures
along with experimental values from Ref.~\cite{kageyama00}. 
%


In order to investigate magnetic anisotropy effects one has to turn to
high-energy resolution calculations with frequency precision
comparable to the magnitude of the DM interaction.  On this frequency
scale we expect to find substancial difference between longitudinal
and transverse components of the spin structure factor. For this
reason we have included the transverse component of the spin structure
factor according to the relation for the differential cross section
\begin{equation}
\frac{{\rm d}^2\sigma}{{\rm d}\Omega\, {\rm d}\omega}\propto\sum_{\mu\nu}
\left(\delta_{\mu\nu}-\frac{q_\mu q_\nu}{q^2}\right)
S_{\mu\nu}({\bf q},\omega).\label{diff}
\end{equation}
For a given direction of the neutron momentum transfer, e.g., ${\bf
q}=(-2,0)$ as used to obtain high-resolution INS data presented in
Fig.~\ref{fig1}(c), Eq.~(\ref{diff}) reduces to a sum of the
transverse and longitudinal part, ${{\rm d}^2\sigma}/{{\rm d}\Omega\,
{\rm d}\omega}\propto S_{yy}({\bf q},\omega)+S_{zz}({\bf
q},\omega)$. In Fig.~\ref{fig1}(c) both contributions as well as their
sum (Eq.~(\ref{diff})) are plotted against the high-resolution INS
data.\cite{gaulin} The best fit is achieved for $J=76.8$~K,
$J'=47.6$~K, $D'_z=1.77$~K, and artificial broadening of the Gaussian
form with $\sigma=0.05$~meV. The splitting $\Delta\sim 0.32$~meV
between the outer two modes belonging to $S_z=\pm1$ single-triplet
excitations originates from a finite value $D'_z$ yielding
$4D'_z/\Delta\sim 1.9$ for the renormalization of the bandwidth due to
quantum fluctuations. This is as well in agreement with the estimate
of Cepas {\it et al.}. \cite{cepas01}

For comparison we also present $S_{zz}({\bf q},\omega)$ of the
simplistic DIM model with $J=\Delta=34~{\rm K}$, where $J$ is chosen
in such a way that SHS and DIM model share identical energy gaps
between the singlet ground state and the excited triplet state.  In
the latter case analytical expression for $S_{zz}({\bf q},\omega)$ can
be straightforwardly derived
\begin{eqnarray}
S_{zz}({\bf q},\omega)\!\! &=&\!\! \pi A({\bf q})\left(\delta (\omega-J)+
e^{-\beta J}\delta (\omega+J)\right)\nonumber\\
&\,+&\! 2\pi B({\bf q})e^{-\beta J}\delta (\omega)\label{sqdim}
\end{eqnarray}
with $A({\bf q})$ and $B({\bf q}) $ given by
%
%
\begin{eqnarray}
A({\bf q})\!\!&=&\!\! \frac{\sin^2\eta (q_x-q_y)+\sin^2\eta (q_x+q_y)}
{4(1+3e^{-\beta J})},\label{aq}\\
B({\bf q})\!\!&=&\!\! \frac{\cos^2\eta (q_x-q_y)+\cos^2\eta (q_x+q_y)}
{4(1+3e^{-\beta J})},\label{bq}
\end{eqnarray}
and $\eta = 0.717$. [Note also that in the limit $T\to \infty$
$\int_{-\infty}^\infty S_{zz}({\bf q},\omega)\, {\rm d}\omega = \pi/4$.] 
At $T=0$ $S_{zz}({\bf q},\omega)$ consists of a single delta function at
$\omega = J$ weighted by $A({\bf q})$.\cite{kageyama00} This peak
corresponds to peak I in the SHS model, while peaks II and III do not
have their counterparts in $S_{zz}({\bf q},\omega)$ of the simplistic DIM
model. With increasing $T$ peaks at $\omega = -J$ and $\omega=0$
appear, weighted by $\pi A({\bf q})\exp(-\beta J)$ and $2\pi B({\bf
q})\exp(-\beta J)$, respectively.

\begin{figure}[htb]
\includegraphics[angle=0,width=8cm,scale=1.0]{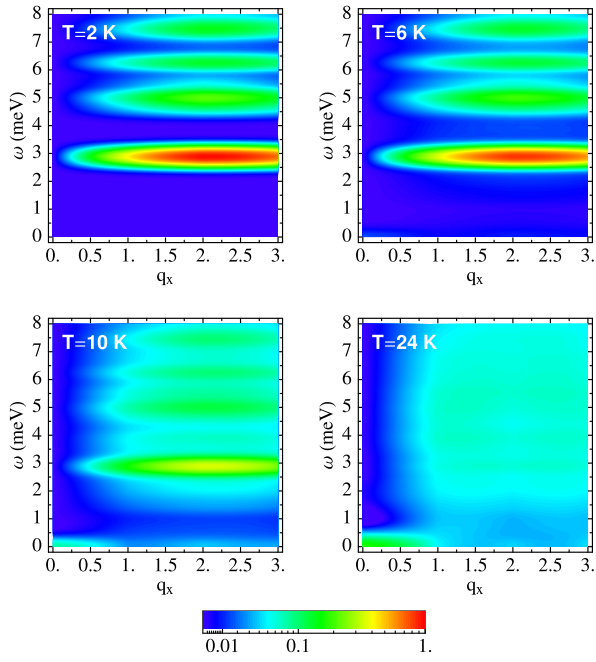}
\vspace{0.4cm} \caption{(Color online) Intensity plot of the spin
structure factor $S_{zz}({\bf q},\omega)$ in the $q_x-\omega$ plane
for different values of $T$ as indicated in the figures. The intensity
scale is logarithmic and $\sigma=0.2$~meV is used to broaden the
spectra as a function of frequency $\omega$.
Note that due to the small system size only integer values of $q_x$
were numerically accesible.}
\label{fig2}
\end{figure}

In Fig.~\ref{fig2} we present a map of $S_{zz}({\bf q},\omega)$ for
different temperatures. Peak I with almost no dispersion is clearly
visible at $\omega \sim 3$~meV with its highest intensity located near
$q_x\sim 2$. Peak II, located at $\omega \sim 5$~meV is also visible
and similarly shows little dispersion. Its intensity is as well
maximal near $ q_x \sim 2$. Note that due to a small system size
$S_{zz}({\bf q},\omega)$ plots were calculated at only a few discrete
values of $q_x$, i.e., $q_x=0.0,1.0,2.0$ and 3.0.
The geometry of the tilted square lattice with $N=20$ sites excludes
half-integer values of $q_x$. This fact prevents us to directly
compare our intensity plot results for the spin structure factor with
the ones shown in Ref.~\cite{gaulin}, where the dispersion of the
lowest triplet mode, attributed mainly to the transverse part of the
spin structure factor, is clearly seen. We have calculated the
transverse component $S_{yy}({\bf q},\omega)$ but since it does not
differ considerably from $S_{zz}({\bf q},\omega)$ on a given energy
scale we do not present it in Fig.~\ref{fig2}.
A final map, shown in Fig.~\ref{fig2}, was obtained by
interpolation between given
integer values of
$q_x$ points. These results are roughly consistent with measurements
by Gaulin {\it et al.}.\cite{gaulin} With increasing $T$ peaks I and
II rapidly decrease (more quantitative analysis of the temperature
dependence follows in the next subsection), while visible response due
to elastic transitions among identical multiplets starts developing
around $\omega =0$ and $q_x=0$.

\subsection{Normalized Peak Intensities}
\begin{figure}[htb]
\includegraphics[angle=0,width=8.5cm,scale=1.0]{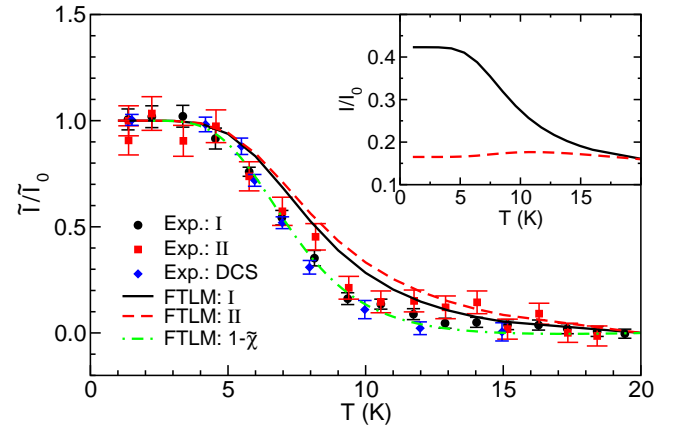}
\vspace{0.4cm} \caption{(Color online) Normalized peak intensities
$\tilde I/\tilde I_0$ of peaks I and II vs. $T$ from Ref.\cite{gaulin}
(filled symbols), FTLM results (lines).  The inset: relative
integrated intensities $I/I_0$ of peaks I and II. \label{fig3}}
\end{figure}

With the purpose to further quantify the agreement of our calculations
with the experiment we present in Fig.~\ref{fig3} the normalized peak
intensities $\tilde I/\tilde I_0$ of the two peaks (I and II) as
functions of temperature along with the measured data taken from
Ref.~\cite{gaulin}. To avoid contributions from the background at
higher temperatures, peak intensities were measured from their values
at $T=20$~K,
\begin{equation}
\tilde I/\tilde I_0=\frac{S_{zz}({\bf q},\omega_P,T)-
S_{zz}({\bf q},\omega_P,20~{\rm K})}{S_{zz}({\bf q},\omega_P,0)-
S_{zz}({\bf q},\omega_P,20~{\rm K})},
\end{equation}
with $\omega_P=3.0$~meV and $\omega_P=5.0$~meV for peak I and II,
respectively. Gaussian broadening with $\sigma=0.4$ meV was used to
obtain peak values of $S_{zz}({\bf q},\omega_P,T)$.
A similar temperature behavior is observed as in INS
measurements,\cite{gaulin,kageyama00} manifesting itself in a rapid
decrease of both peak intensities with temperature for $T$ far below
the gap value $\Delta=34$~K.
%
%
Taking this fact into account, the agreement between experimental
values and numerical calculations of $\tilde I/\tilde I_0$ is
reasonable even though not ideal.  However, as already suggested by
Gaulin {\it et al.},\cite{gaulin} nearly perfect agreement between
experiment and rescaled complement of the uniform static
susceptibility $1-\tilde \chi = 1-\chi_0(T)/\chi_0(T=20~{\rm K})$ is
found where $\chi_0 = \langle{S^z_{tot}}^2\rangle/NT$, and $S^z_{tot}$
represents the $z$-component of the total spin. We were unable to find
a direct analytical connection between the two quantities, i.e.,
$1-\tilde \chi$ and $\tilde I/\tilde I_0$, on the other hand the
difference between the above mentioned quantities in numerical results
are obvious from Fig.~\ref{fig3}.  Furthermore, analytical
calculations of $1-\tilde \chi$ and $\tilde I/\tilde I_0$ on DIM model
also indicate a different $T$ dependence. This leads us to the
conclusion, that nearly perfect agreement with $1-\tilde\chi$ and
experimental results of Ref.~\cite{gaulin} may be accidental.

In the inset of Fig.~\ref{fig3} we show the integrated intensities of
the two peaks defined as $I(\omega_1,\omega_2)/I_0 =
\int_{\omega_1}^{\omega_2} {\rm d}\omega S_{zz}({\bf q},\omega)/\langle
S_{\bf q}S_{-\bf q}\rangle$ where the limits of integration defining
integrated peak intensities are defined as follows: $I_I=I(2~{\rm
meV},4~{\rm meV})$ and $I_{II}=I(4~{\rm meV},6~{\rm meV})$
for peaks I and II, respectively. We observe a distinctive
difference in temperature behavior between $I_I/I_0$ on the one
hand and $I_{II}/I_0$ on the other. While $I_I/I_0$ substantially
decreases with increasing temperature similarly to $\tilde
I/\tilde I_0$, indicating on a considerable shift of the spectral
weight away from transition I, $I_{II}/I_0$ shows even a slight
temperature increase. We suggest that this difference is caused by
a different nature of the transition from the ground state to the
localized triplet (peak I) in contrast to transitions to states
near or else within continuum (peak II). This behavior is as well
in agreement with INS measurements \cite{kageyama00} that show
peak I being only resolution limited while peaks II and III show
intrinsic linewidths.

\begin{figure}[htb]
\includegraphics[angle=0,width=7.2cm,scale=1.0]{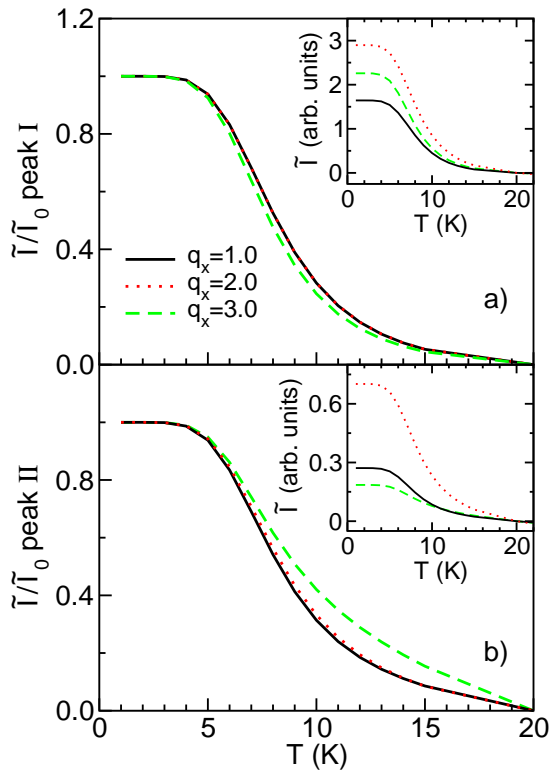}
\vspace{0.4cm} \caption{(Color online) Normalized peak intensities
$\tilde I/\tilde I_0$ of peaks (a) I and (b) II vs. $T$ calculated at
different values of ${\bf q}=(q_x,0)$. In insets unrenormalized peak
intensities measured from the peak intensity position at $T=20$~K are
shown. \label{fig4}}

\end{figure}

We now explore the ${\bf q}$-dependence of peak intensities. In
Fig.~\ref{fig4} we present normalized values of peak intensities
vs. $T$ for various values of $q_x$ at fixed $q_y=0$ and find
nearly perfect scaling of $\tilde I/\tilde I_0$ for peak I,
Fig.~\ref{fig4}(a), while scaling breaks down at $q_x=3.0$ for
peak II, Fig.~\ref{fig4}(b). Such behavior is characteristic also
for the simpler DIM model that possesses a single temperature
scale $J$. This result suggests that a single temperature scale is
responsible for the $T$ dependence of peak I for all different
values of ${\bf q}$.  In the insets of Figs.~\ref{fig4}(a) and
\ref{fig4}(b) we present absolute values of peak intensities for
different values of $q_x$. Intensities of both peaks I and II
reach their maximum values at low-$T$ near $q_x\sim2.0$. Taking
into account our rather poor resolution in the $q_x$-space, we
find these results to be roughly consistent with recent
high-resolution INS measurements by Gaulin {\it et
al.}.\cite{gaulin}

\subsection{Thermodynamic Properties}
\begin{figure}[htb]
\includegraphics[angle=0,width=7.2cm,scale=1.0]{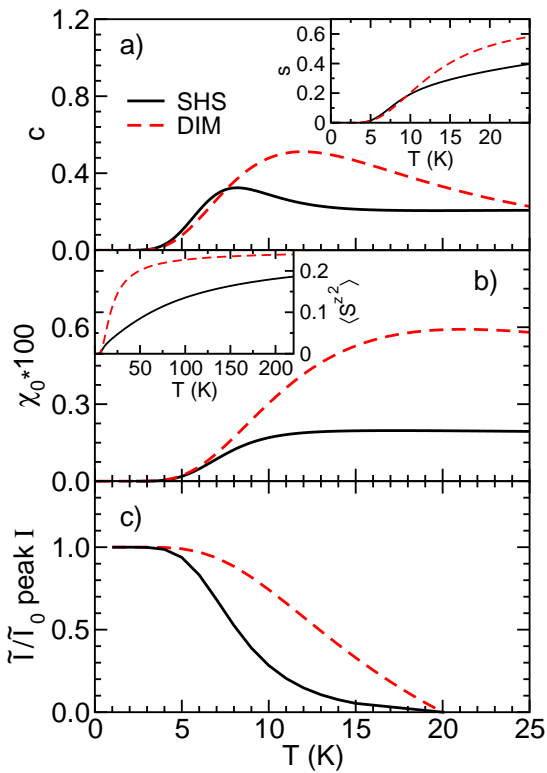}
\vspace{0.4cm} \caption{(Color online) Thermodynamic properties of the
SHS model (full lines) and DIM model with $J=34~{\rm K}$ (dashed
lines): (a) specific heat $c$ (entropy $s$ in the inset) vs. $T$ and
(b) uniform static susceptibility $\chi_0$ vs. $T$. For completeness
and to facilitate comparison we present in (c) normalized peak
intensity of peak I $\tilde I/\tilde I_0$ vs. $T$ measured from values
at $T=20$~K as shown in Fig. \ref{fig3}.
\label{fig5}}
\end{figure}

Next, we will connect the spectral data with thermodynamic properties
of the Hamiltonian defined in Eq.~(\ref{Hamil}). For this reason we
present in Fig.~\ref{fig5}(a) the specific heat (per spin)
$c=T(\partial s/\partial T)=(\langle H^{2}\rangle-\langle
H\rangle^{2})/NT^{2}$ and the entropy density $s= lnZ/N+\langle
H\rangle/NT$, where $Z$ is the statistical sum and $N$ is the number
of spins in the system. Specific heat peaks around $T=T_{max}\sim
8$~K, where we also observe a rapid drop of $\tilde I/\tilde I_0$ [see
Fig.~\ref{fig3}], furthermore, this temperature also coincides with
the steepest ascent of $s$. The peak in $c$, located well below the
value of the gap, $T_{max}\sim 0.24 \Delta$, is a consequence of
excitations from the ground state to localized singlet and triplet
states. This peak is followed by a broad shoulder above $T > 15$~K
which is due to excitations in the continuum.
Note that $c$, obtained using the same method and for slightly
different values of $J,J'$ and ${\bf D}$, apart for additional DM
terms, fits measured specific heat data of SrCu$_{2}$(BO$_{3}$)$_{2}$
in a wide range of applied external magnetic fields.\cite{jorge03}
For comparison we also present $c$ and $s$ of the DIM model with
$J=\Delta=34~{\rm K}$ that can be solved analytically. We should point
out that even in a simple DIM model $c$ peaks well below the gap
value, i.e., $T_{max}=0.35
\Delta\sim 11.9$~K.

In Fig.~\ref{fig5}(b) we present the uniform static spin
susceptibility, $\chi_0 = \langle{S^z_{tot}}^2\rangle/NT$, where
$S^z_{tot}$ represents the $z$-component of the total spin. While
comparison of $\chi_0$ with experimental data
%
%
was presented elsewhere, \cite{jorge03} in this work we present it
along with other thermodynamic properties just to gain a more complete
physical picture of the temperature dependence of $S_{zz}({\bf
q},\omega)$. The steepest increase in $\chi_0$ vs. $T$ coincides with
the peak in $c$ and, at least approximately, with the steepest
decreases of $\tilde I/\tilde I_0$, presented in Fig.~\ref{fig3}. Due
to the existence of the spin gap, $\chi_0(T\lsim 5~{\rm K})\sim 0$,
where the temperature interval $T\lsim 5$~K in turn corresponds to the
plateau of $\tilde I/\tilde I_0$ seen in experimental results of peaks
I and II as well as in our numerical simulations.

We would like to make some general remarks on comparing thermodynamic
properties of the SHS model and the DIM model with identical gaps
between the ground state and first excited states. Such a direct
comparison may assist in understanding the influence of spin
frustration and the proximity of gapless excitations in the SHS model
on its thermodynamic properties. In particular, the specific heat $c$
of the SHS model peaks at lower temperature than $c$ of the DIM model
and shows two maxima in contrast to a single, Schottky-like maximum
seen in DIM model. The entropy of SHS model reveals slower increase
with $T$ than that of the DIM model. And finally, the peak value of
$\chi_0$ is almost three times lower in the SHS model than in the DIM
model which in turn implies that spin fluctuations,
$\langle{S^z}^2\rangle$, [see the inset of Fig.~\ref{fig5}(b)] of the
SHS model are suppressed in comparison to the DIM model.

\subsection{Static Spin Susceptibility $\chi({\bf q})$}

\begin{figure}[htb]
\includegraphics[angle=0,width=8.5cm,scale=1.0]{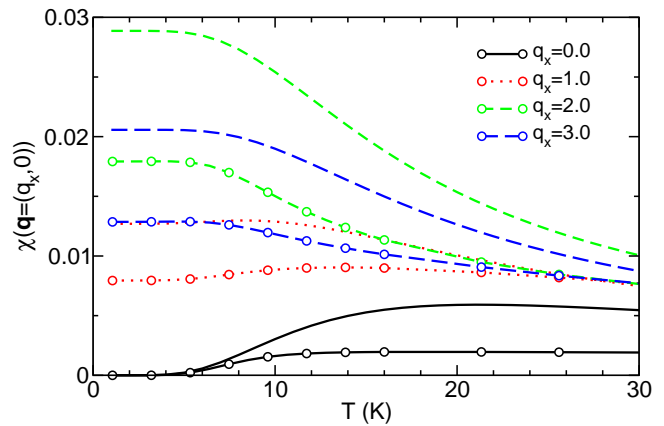}
\caption{(Color online) $\chi({\bf q}=(q_x,0))$ vs. $T$ for several
values of $q_x$. Circles connected with lines present results of the
SHS model, lines present results of the DIM model, Eq.~(\ref{chiqdim})
with $J=34$~K. \label{fig6}}
\end{figure}

Finally, we present in Fig.~\ref{fig6} the static spin susceptibility
\begin{eqnarray}
\chi({\bf q})\!\! &=&\!\! \frac{1}{\pi}{\cal P}\int\limits_{-\infty}^{\infty}\! 
\frac{\chi^{''}{(\bf q,\omega)}}{\omega}\ {\rm d}\omega,\label{chiq} \\
\chi^{''}(\bf q,\omega)\!\! &=&\!\! \left(1-e^{-\beta
\omega}\right)S_{zz}({\bf q},\omega), \label{chippq}
\end{eqnarray}
as a function of ${\bf q}=(q_x,0)$. Besides fulfilling theoretical
interest, $\chi({\bf q})$ can also be used to compute, e.g., spin-spin
nuclear relaxation rate $1/T_2$. Along $\chi({\bf q})$ of the SHS
model we present for comparison results for the DIM model where
analytical result can be readily obtained using Eqs.~(\ref{sqdim}),
(\ref{chiq}), and (\ref{chippq}):
\begin{equation}
\chi({\bf q})=2A({\bf q})\left(1-e^{-\beta  J}\right)/J+2B({\bf
q}) e^{-\beta J}\beta\label{chiqdim}
\end{equation}
where $A({\bf q})$ and $B({\bf q})$ are defined in Eqs.~(\ref{aq}) and
(\ref{bq}).
At low-$T$, i.e., $T \lsim 5$~K, $\chi({\bf q})$ vs. $T$ is nearly
$T$-independent which is a consequence of the spin gap. As a function
of $q_x$ it reaches its maximum value near $q_x\sim 2.0$ in accord
with the prediction of the DIM model result,
Eq.~(\ref{chiqdim}). Observed $T$-dependence is again similar to the
DIM model prediction. At higher temperatures, i.e., $T\gsim 300$~K,
$\chi({\bf q})$ merges with universal, ${\bf q}$-independent form,
i.e., $\chi({\bf q})=1/4T$.

\section{Conclusions}

In conclusion, we have computed dynamical spin structure factor at
finite temperatures. Frequency dependence of $S_{zz}({\bf q},\omega)$
at $T=2$ and 24 K agree reasonably well with INS measurements
\cite{kageyama00} on a large energy scale despite rather poor
frequency resolution caused by finite-size effects.
High-resolution data for the lowest triplet excitation
\cite{gaulin}, on the other hand, is almost perfectly captured by
calculated transverse and longitudinal components, $S_{yy}({\bf
q},\omega)+S_{zz}({\bf q},\omega)$, showing the influence of
anisotropy present in the system.
Comparison of results obtained on systems with $N=16$ and 20 sites
reveals that the positions of peaks I and II are reasonably well
reproduced on $N=20$ system, while the peak III position and its
structure are less accurate.
We should also note that although in a 16-site system finite-size
effects are somewhat more pronounced, this cluster enables one to
calculate half-integer values of $q_x$ as well. By fitting the data
from Ref.~\cite{gaulin} for all measured $q_x$ on a system with $N=16$
sites, we have concluded that additional NN in-plane as well as
'forbiden' NN out-of-plane DM interactions are required to succesfully
explain the structure and dispersion of peak I. However, the best fit
is obtained with slightly different set of parameters as used in this
work.

Temperature dependence of normalized peak intensities $\tilde I/\tilde
I_0$ agrees well with INS measurements.\cite{kageyama00,gaulin} Our
calculations predict that $\tilde I/\tilde I_0$ of peak I should be
${\bf q}$-independent. Such behavior is in agreement with the DIM
model prediction for peak I, while peak II is anyhow absent in this
simplistic model. Our results are thus consistent with a proposition
that a single temperature scale is responsible for the $T$ dependence
of peak I for all different values of ${\bf q}$.  This statement does
not take into account a possible small dispersion of peak I due to DM
interaction or (and) due to high-order processes in
$J'/J$. \cite{miyahara03} From comparison of temperature dependence of
$\tilde I/\tilde I_0$ with thermodynamic properties it is obvious that
strong $T$-dependence of $\tilde I/\tilde I_0$, occurring well below
the value of the spin gap, is in accord with strong $T$-dependence of
other thermodynamic properties. The temperature of the steepest
decrease of $\tilde I/\tilde I_0$ coincides with the peak in $c$ and
the steepest increase of $s$ as well as of $\chi_0$.

There is obviously a need for further, less finite-size dependent
calculations that will clarify many unanswered questions as are, e.g.,
the role of DM terms in explaining small dispersion of peaks I and II
observed in high-resolution INS experiments \cite{gaulin}, an
explanation of unusual temperature dependence of ESR lines
\cite{nojiri03} that seem to decrease in width as the temperature
increases, the occurrence of magnetization plateaus, etc.
Nevertheless, the main features of temperature-dependent dynamic
properties of the SHS model seem to be well captured by the FTLM on
small lattices which is in turn reflected in a good agreement with
experiments.

\acknowledgments  We acknowledge inspiring discussions with
C.D. Batista and B.D. Gaulin that took place during the preparation of
this manuscript. We also acknowledge the financial support under
contract P1-0044.

\end{document}